\documentclass{article}
\usepackage{amssymb,amsmath,cite,bm}
\usepackage{graphicx}
\def\rd{\mathrm{d}}
\begin{document}

\begin{titlepage}
\begin{center}
{\large\textbf{A suspended rope wrapped around a cylinder}}

\vspace{2\baselineskip}
{\sffamily Mohammad~Khorrami~\footnote{e-mail: mamwad@mailaps.org},
Amir~Aghamohammadi~\footnote{e-mail: mohamadi@alzahra.ac.ir}.}

\vspace{2\baselineskip}
\textit{Department of Physics, Alzahra University, Tehran 19938-93973, IRAN}
\end{center}


\begin{abstract}
\noindent The problem of a suspended rope wrapped around
a fixed cylinder is studied. If the suspension
force is larger than a certain threshold (which is larger than
the weight of the rope), the rope would remain tightly wrapped around
the cylinder. For suspension forces smaller than that threshold but larger
than another threshold, the rope becomes loose (loses contact with
the cylinder) at some points, but still remains at rest. Theses
thresholds are obtained when there is no friction. The system is then
analyzed also with friction, and the threshold of tight-wrapping is obtained
for that case as well.
\end{abstract}
\end{titlepage}
\newpage
\section{Introduction}
A standard problem in elementary mechanics is the study of the equilibrium
of a particle on a surface. If the particle moves on the surface, there
could be a point where the normal force changes direction. As a nonadhesive
surface cannot produce a normal force which pulls the particle toward
the surface, at this point the normal force becomes zero and the particle
loses contact with the surface. Finding the point where a particle sliding
on a frictionless sphere loses contact with the sphere has been studied in many
books on elementary mechanics (\cite{Hall, KLep,Irodov,Morin} for example).
In \cite{Amireur}, this problem was extended to arbitrary surfaces, and
conditions were found for the point where the particle leaves the surface.

The problem could be more complex if the moving object is not a point particle,
for example when it is one dimensional (such as a rope). A simple example
of this, is the problem of a rope wrapped around a pole, where due to friction,
a small force in one side can keep the rope from sliding. Some other examples
regarding flexible chains and strings have been considered in \cite{Ram}.
In \cite{Bay} the case of a rope wrapped around a circular post has been studied,
where the rope crosses over itself several times.
The equilibrium of a rope wrapped  around a solid body, and the problem of
two ropes crossing each other lying on a surface, have been
investigated in \cite{Madd}.

The problem addressed here is that a massive uniform rope is wrapped
around a fixed cylinder, so that the rope has two vertical ends 
supported by external forces. The system is in a uniform 
gravitational field. It could be that there is a friction between 
the rope and the cylinder, or not. Both cases are studied. It could be 
that the rope is tightly wrapped around the cylinder, as in figure 
\ref{fig1}, or that the rope is not touching the bottom of the cylinder, 
as in figure \ref{fig2}. The questions are the following.
\begin{itemize}
\item What is the minimum value of the supporting force, in order that
the rope is tightly wrapped around the cylinder?
\item If the supporting force is less that the above value, how much
does the rope drop for a given supporting force?
\item What is the minimum value of the supporting force, in order that
the rope does not completely fall?
\end{itemize}

In section 2, conditions are investigated for the rope to remain tightly
wrapped around the cylinder. It is shown that there is a minimum value 
for the external force holding the rope, in order that the rope remains 
tightly wrapped around the cylinder, and that minimum is determined. 
In section 3, a situation is studied where the rope has lost contact
with some of the bottom part of the cylinder. When this happens,
the ends of the rope, where the supporting force is applied, fall
by a certain amount $S$. It is shown that two constraints are to be 
satisfied, one is that the normal force density applied to the rope 
by the cylinder should be outward, and the other is that the loose part 
of the rope should be supported by the tension of the rope. It is further 
shown that this second constraint is stronger. The angle at which the rope 
looses contact with the cylinder is found, in terms of 
the supporting force. It is shown that there is a minimum value for 
the supporting force in order to prevent the rope from falling indefinitely. 
Also, it is seen that the relation of the angle of contact-loss and 
the supporting force consists of a stable part and an instable part. 
In section 4, the effect of friction is taken into account and conditions 
are found that the rope be at the threshold of loosing contact with 
the cylinder. Section 5 is devoted to the concluding remarks.
\section{The tight rope}
\begin{figure}
\begin{center}
\includegraphics[scale=0.7]{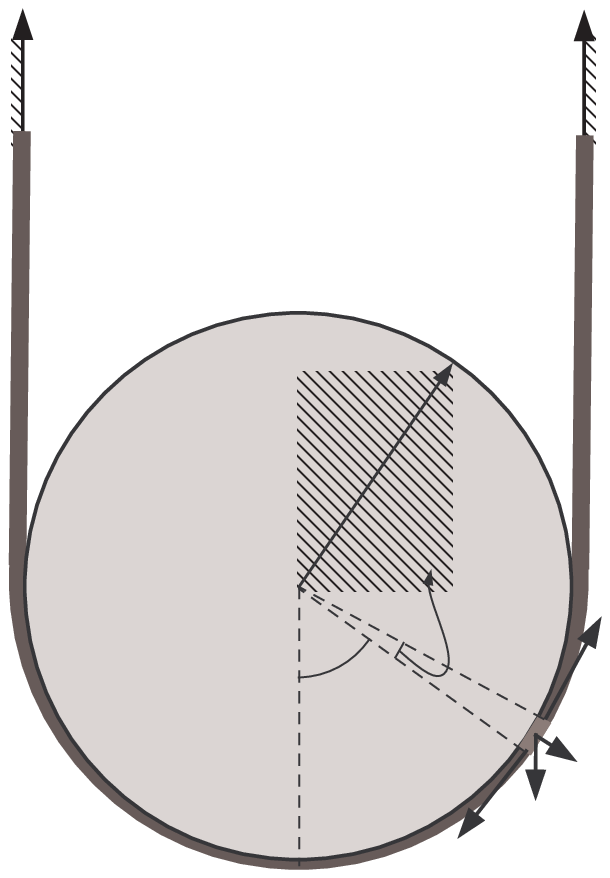}
\setlength{\unitlength}{1pt}
\put(-8,60){A}
\put(-43,3){$T$}
\put(-28,10){$\delta m\, g$}
\put(-10,25){$\delta N$}
\put(-5,50){$T+\delta T$}
\put(-130,185){$F_0$}
\put(-14,185){$F_0$}
\put(-63,82){$R$}
\put(-60,38){$\theta$}
\put(-45,68){$\delta\theta$}
\caption{\label{fig1}}{a rope wrapped around a cylinder}
\end{center}
\end{figure}

Consider a uniform rope of linear mass density $\lambda$, and
length $L$. The rope is wrapped around a stationary cylinder of
radius $R$, is in a vertical plane, and each of its ends are pulled
upward with the force $F$, as in figure \ref{fig1}. An element of
the rope which is in contact with the cylinder, spanning an arc angle
of $\delta\theta$, has the mass $\delta m$:
\begin{equation}\label{01}
\delta m=R\,\lambda\,\delta\theta.
\end{equation}
The linear weight density is denoted by $w$:
\begin{equation}\label{02}
w=\lambda\,g,
\end{equation}
where $g$ is the acceleration due to gravity. Denoting the tension of
the rope by $T$, and the normal force acting on this element by $\delta N$,
the Newton's equations for this element (at rest), projected on the tangent 
and the normal, are
\begin{align}\label{03}
&(T+\delta T)\,\cos\frac{\delta\theta}{2}-T\,\cos\frac{\delta\theta}{2}-(w\,R\,\delta\theta)\,
\sin\theta=0.\\ \label{04}
&\delta N-(T+\delta T)\,\sin\frac{\delta\theta}{2}-T\,\sin\frac{\delta\theta}{2}+
(w\,R\,\delta\theta)\,\cos\theta=0.
\end{align}
Dividing the above equations by $\delta\theta$, and sending
$\delta\theta$ to 0, one arrives at a differential equation for
$T$ and an algebraic equation for $\delta N/\delta\theta$.
It would be convenient to introduce the new parameters $\tau$
and $n$ through
\begin{align}\label{05}
\tau&:=\frac{T}{w\,R}.\\ \label{06}
n&:=\frac{1}{w\,R}\,\frac{\delta N}{\delta\theta}.
\end{align}
Basically, $\tau$ is a dimensionless tension and $n$ is
a dimensionless angular density of the normal force.
In terms of these, equations (\ref{03}, \ref{04}) become,
\begin{align}\label{07}
\frac{\rd\tau}{\rd\theta}&=\sin\theta.\\ \label{08}
n&=\tau-\cos\theta.
\end{align}
These result in
\begin{align}\label{09}
\tau(\theta)&=\tau\left(\frac{\pi}{2}\right)-\cos\theta.\\ \label{10}
n(\theta)&=\tau\left(\frac{\pi}{2}\right)-2\,\cos\theta.
\end{align}
$\tau(\pi/2)$ is the dimensionless tension at the point A.
The normal force density should be nonnegative, resulting in
\begin{equation}\label{11}
\tau\left(\frac{\pi}{2}\right)\geq 2.
\end{equation}
This gives a criterion for $F$:
\begin{align}\label{12}
F\geq&\,F_0,\nonumber\\
F_0:=&\,\frac{w\,L}{2}+w\,R\,\left(2-\frac{\pi}{2}\right).
\end{align}
If the external force $F$ is less than $F_0$, then the rope would
lose contact with the cylinder at some bottom part of the cylinder.
One notes that this force is larger than half the weight of the rope:
$(2\,F_0)$ should cancel the weight of the rope as well as the downward
component of the normal force.
\section{The loose rope}
\begin{figure}
\begin{center}
\includegraphics[scale=0.7]{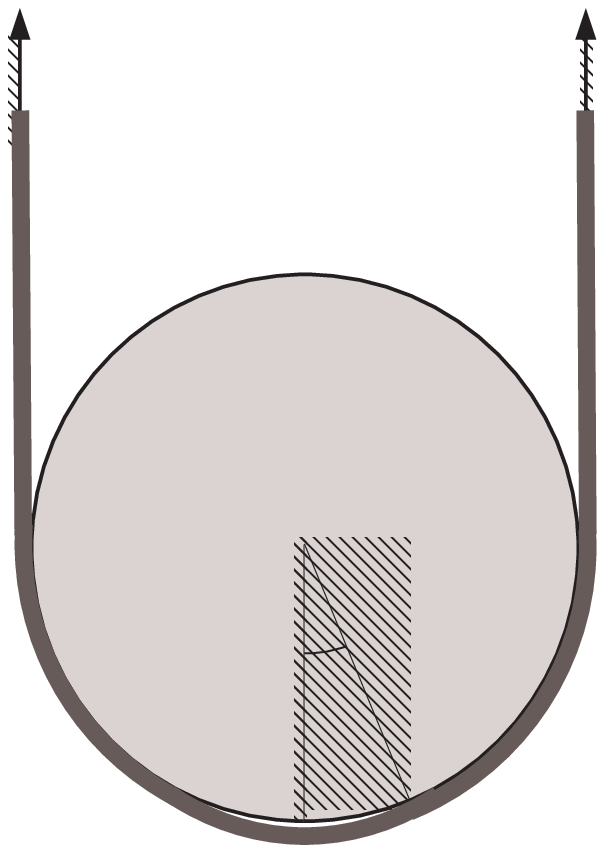}
\setlength{\unitlength}{1pt}
\put(-110,12){B$'$}
\put(-34,12){B}
\put(-5,60){A}
\put(-62,35){$\theta_0$}
\caption{\label{fig2}}{a rope loosely wrapped around a cylinder}
\end{center}
\end{figure}

Suppose the external force $F$ is less than $F_0$.
Then the rope would lose contact with the cylinder till
some angle $\theta_0$, figure \ref{fig2}. Equations
(\ref{05}) through (\ref{10}) still hold for larger angles.
The condition that $n$ vanishes at $\theta_0$, gives the tension
at the angle $\theta_0$ through
\begin{equation}\label{13}
\tau_1(\theta_0)=\cos\theta_0.
\end{equation}
But at least for large $\theta_0$, this is obviously wrong,
as it says that  $\tau_1(\pi/2)$ tends to $0$ for $\theta_0$
tending to $(\pi/2)$. What would then cancel the weight of the rope?

\begin{figure}
\begin{center}
\includegraphics[scale=0.7]{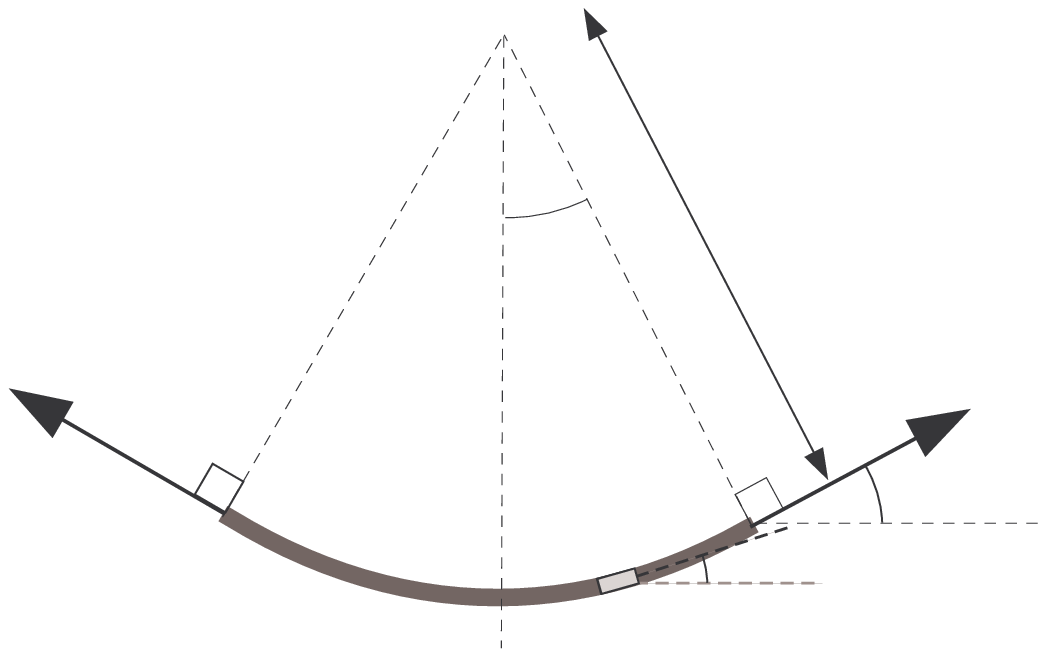}
\setlength{\unitlength}{1pt}
\put(-110,85){$\theta_0$}
\put(-35,35){$\theta_0$}
\put(-65,22){$\beta$}
\put(-67,83){$R$}
\caption{\label{fig3}}{a freely hanging rope}
\end{center}
\end{figure}

The problem is that there is another constraint as well,
apart from the fact that the normal force should be nonnegative,
and that constraint is stronger.
The other constraint which should be taken into account, comes
from that part of the rope which is not in contact with the cylinder.
In that part, for an element of the length $\delta s$ of the rope, (see figure \ref{fig3}) one has
\begin{align}\label{14}
\delta (T\,\cos\beta)&=0,\\ \label{15}
\delta (T\,\sin\beta)&=w\,\delta s,
\end{align}
which gives
\begin{align}\label{16}
\tau\,\cos\beta&=c,\\ \label{17}
\frac{\rd(\tau\,\sin\beta)}{\rd x}&=\frac{1}{R\,\cos\beta},
\end{align}
where $c$ is a constant, $\tan\beta$ is the slope of the hanging part of the rope,
$x$ is the horizontal coordinate, and $y$ is the vertical coordinate (upward), so
\begin{equation}\label{18}
\tan\beta=\frac{\rd y}{\rd x}.
\end{equation}
One then has
\begin{equation}\label{19}
c\,R\,\frac{\rd^2 y}{\rd x^2}=\left[1+\left(\frac{\rd y}{\rd x}\right)^2\right]^{1/2},
\end{equation}
leading to
\begin{equation}\label{20}
\frac{\rd y}{\rd x}=\sinh\left(\frac{x}{c\,R}\right),
\end{equation}
where $x=0$ has been taken the point at which the rope is horizontal. So,
\begin{equation}\label{21}
y=c\,R\,\cosh\left(\frac{x}{c\,R}\right),
\end{equation}
with an additive constant appropriately chosen. This is
the well-known hanging-rope equation. Denoting by $\ell$ half
the length of the hanging part,
\begin{align}\label{22}
\ell&=\int_0^{R\,\sin\theta_0}\rd x\,\left[1+\left(\frac{\rd y}{\rd x}\right)^2\right]^{1/2},
\nonumber\\
&=c\,R\,\sinh\left(\frac{\sin\theta_0}{c}\right),
\end{align}
the vertical component of the equilibrium equation for the hanging part of the rope is
\begin{equation}\label{23}
\frac{\ell}{R}=\tau(\theta_0)\,\sin\theta_0.
\end{equation}
Combining this with (\ref{16}), results in
\begin{equation}\label{24}
\frac{\ell}{c\,R}=\tan\theta_0,
\end{equation}
so that
\begin{equation}\label{25}
\tan \theta_0=\sinh\left(\frac{\sin\theta_0}{c}\right).
\end{equation}
Hence
\begin{align}\label{26}
c&=\frac{\sin\theta_0}{\sinh^{-1}(\tan\theta_0)}.\\ \label{27}
\frac{\ell}{R}&=\frac{\sin\theta_0\,\tan\theta_0}{\sinh^{-1}(\tan\theta_0)}.\\ \label{28}
\tau_2(\theta_0)&=\frac{\tan\theta_0}{\sinh^{-1}(\tan\theta_0)}.
\end{align}
Comparing equations (\ref{13}) and (\ref{26}), one notes that 
$\tau_1(\theta_0)$ is always smaller than 1, and $\tau_2(\theta_0)$ 
is always larger than 1, showing that $\tau_2(\theta_0)$ is
larger than $\tau_1(\theta_0)$, hence the constraint on tension 
due to the hanging part of the rope is stronger than the constraint 
due to the normal force being nonnegative.

Regarding the force $F$, one has
\begin{align}\label{29}
F&=(w\,R)\,\tau\left(\frac{\pi}{2}\right)+
w\,\left[\frac{L}{2}-\ell-\left(\frac{\pi}{2}-\theta_0\right)\,R\right],\nonumber\\
&=(w\,R)\,[\tau(\theta_0)+\cos\theta_0]+
w\,\left[\frac{L}{2}-\ell-\left(\frac{\pi}{2}-\theta_0\right)\,R\right],\nonumber\\
&=F_0+w\,R\,\left[\tau(\theta_0)+\cos\theta_0+\theta_0-2-\frac{\ell}{R}\right],
\end{align}
so that
\begin{equation}\label{30}
\frac{F}{w\,R}=\frac{F_0}{w\,R}+
\left[\frac{(1-\sin\theta_0)\,\tan\theta_0}{\sinh^{-1}(\tan\theta_0)}+\cos\theta_0+\theta_0-2\right].
\end{equation}
Also the tension at the point A would satisfy
\begin{equation}\label{31}
\tau_\mathrm{A}(\theta_0)=\frac{\tan\theta_0}{\sinh^{-1}(\tan\theta_0)}+\cos\theta_0.
\end{equation}
For $\theta_0=0$, the right-hand side is equal to 2, as expected.

If the rope is loosened, so that
\begin{equation}\label{32}
\tau_\mathrm{A}(\theta_0)=2-\varepsilon,
\end{equation}
the ends of the rope (where the supporting force is applied) fall by a certain amount
$S$. For a small value of $\epsilon$, one would have
\begin{align}\label{33}
2-\varepsilon&=\frac{\tan\theta_0}{\sinh^{-1}(\tan\theta_0)}+\cos\theta_0,\\ \label{34}
&=2\,\left(1-\frac{\theta_0^2}{6}\right)+\mathrm{O}(\theta_0^4),
\end{align}
so that
\begin{equation}\label{35}
\theta_0=(3\,\varepsilon)^{1/2}+\mathrm{O}(\varepsilon^{3/2}).
\end{equation}
For $S$, one has
regarding $\ell$,
\begin{align}\label{36}
S&=\ell-R\,\theta_0,\nonumber\\
&=R\left[\frac{\sin\theta_0\,\tan\theta_0}{\sinh^{-1}(\tan\theta_0)}-\theta_0\right],\nonumber\\
&=R\left[\frac{2\,\theta_0^5}{45}+\mathrm{O}(\theta_0^7)\right],
\end{align}
so that
\begin{equation}\label{37}
S=R\left[\frac{2}{5}\,(3\,\varepsilon^5)^{1/2}+\mathrm{O}(\varepsilon^{7/2})\right].
\end{equation}

\begin{figure}
\begin{center}
\includegraphics[scale=0.4]{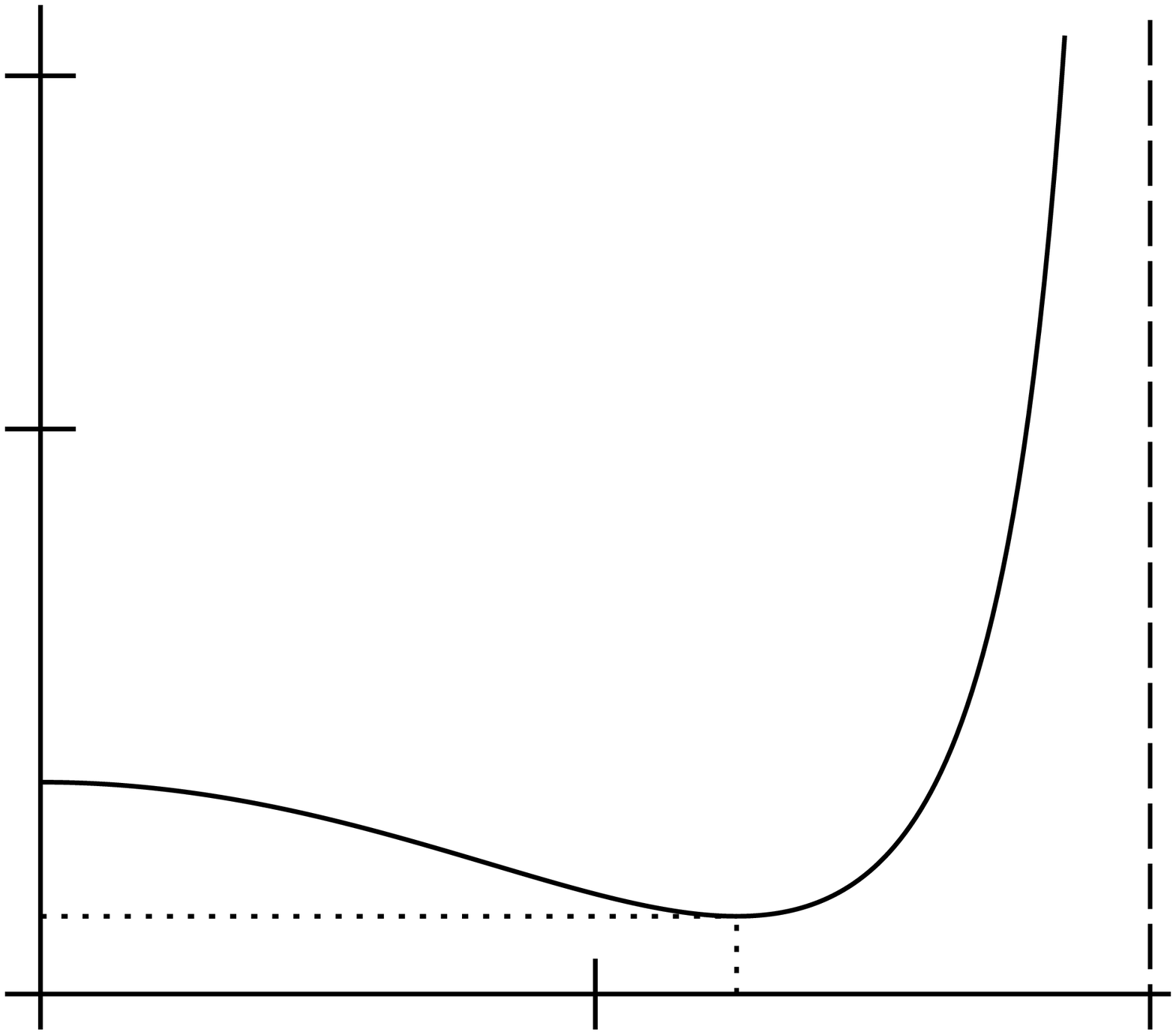}
\setlength{\unitlength}{1pt}
\put(0,21){$\theta_0$}
\put(-15,5){$\pi/2$}
\put(-87,5){$\theta_\mathrm{m}$}
\put(-115,5){$\pi/4$}
\put(-215,5){$0$}
\put(-235,21){$1.7$}
\put(-235,38){$\tau_\mathrm{m}$}
\put(-235,60){$2$}
\put(-235,125){$2.5$}
\put(-235,190){$3$}
\put(-215,210){$\tau_\mathrm{A}$}
\caption{\label{fig4}}{$\tau_\mathrm{A}$ (the dimensionless tension at the point $A$) versus $\theta_0$}
\end{center}
\end{figure}

Inspection of (\ref{31}) shows that (see figure \ref{fig4})
\begin{alignat}{2}\label{38}
&\mbox{$\tau_\mathrm{A}(\theta_0)$ is decreasing with $\theta_0$},&\quad&\theta_0<\theta_\mathrm{m},
\nonumber\\
&\mbox{$\tau_\mathrm{A}(\theta_0)$ is increasing with $\theta_0$},&\quad&\theta_0>\theta_\mathrm{m},
\nonumber\\
&\tau_\mathrm{A}(\theta_0)\to\infty,&\quad&\theta_0\to\frac{\pi}{2},
\end{alignat}
where $\theta_\mathrm{m}$ is the point at which the derivative of
$\theta_\mathrm{A}$ vanishes:
\begin{align}\label{39}
&\sinh^{-1}(\tan\theta_\mathrm{m})-(\sin\theta_\mathrm{m})\,
\{1+(\cos^2\theta_\mathrm{m})\,[\sinh^{-1}(\tan\theta_\mathrm{m})]^2\}=0.\\ \label{40}
&\theta_\mathrm{m}=0.9855\,{\rm rad},\ {\rm or} \ 56.46^\circ.
\end{align}
Denoting $\tau_\mathrm{A}(\theta_\mathrm{m})$ by $\tau_\mathrm{m}$,
\begin{align}
\tau_\mathrm{m}&=\tau_\mathrm{A}(\theta_\mathrm{m}),\nonumber\\
&=1.810,
\end{align}
it is seen that
\begin{itemize}
\item For $\tau_\mathrm{A}<\tau_\mathrm{m}$, there are no solutions for $\theta_0$.
This means that if $\tau_\mathrm{A}<\tau_\mathrm{m}$, the tension cannot
support the rope and the rope falls indefinitely.
\item For $\tau_\mathrm{m}<\tau_\mathrm{A}<2$, there are two solutions for
$\theta_0$. One of these is less than $\theta_\mathrm{m}$, the other
is larger than $\theta_\mathrm{m}$. The first solution corresponds to
a stable equilibrium, which means that changing $\tau_\mathrm{A}$
slightly results to a slight change in $\theta_0$. The second solution
corresponds to an unstable equilibrium: increasing $\tau_\mathrm{A}$
slightly results in $\theta_0$ moving towards the first solution and
a stable equilibrium being established near the first solution, while
decreasing $\tau_\mathrm{A}$ slightly results in the rope falling indefinitely.
\item For $\tau_\mathrm{A}>2$, there is only one solution for
$\theta_0$, and that solution is larger than $\theta_\mathrm{m}$.
The equilibrium is unstable. Increasing $\tau_\mathrm{A}$
slightly results to the rope becoming tight ($\theta_0$ tending to zero), while
decreasing $\tau_\mathrm{A}$ slightly results in the rope falling indefinitely.
\end{itemize}
\section{Rope with the friction}
If the rope is tight, and there is friction, Newton equations change to
\begin{align}\label{42}
&\frac{\rd\tau}{\rd\theta}+f=\sin\theta,\\ \label{43}
&n+\cos\theta=\tau,
\end{align}
where $f$ is the dimensionless angular density of the normalized friction:
\begin{equation}\label{44}
f:=\frac{1}{w\,R}\,\frac{\delta F_\mathrm{fr}}{\delta\theta},
\end{equation}
$\delta F_\mathrm{fr}$ being the friction felt by a part of
the rope of the angular size of $\delta\theta$. Differentiating
the second equation, to eliminate the tension, one arrives at
\begin{equation}\label{45}
\frac{\rd n}{\rd\theta}+f=2\,\sin\theta.
\end{equation}
If there is a point $\theta'$ where $n(\theta')$ vanishes, 
then $f(\theta')$ vanishes as well, so that 
$\rd n/\rd\theta$ is positive at $\theta'$, 
unless $\theta'$ is zero. But if $\rd n/\rd\theta$ 
is positive at $\theta'$, then $n(\theta)$ is negative at some $\theta$ 
with $\theta<\theta'$, which is not acceptable. So $n$ can vanish only 
at $\theta=0$. So relaxing the rope, the first point which loses its 
contact with the cylinder is the lowest point of the rope. Knowing that
\begin{equation}\label{46}
f(\theta)\geq\mu\,n(\theta),
\end{equation}
it is seen that at the beginning of sliding,
\begin{align}\label{47}
n_3(0)&=0.\\ \label{48}
f_3(\theta)&=\mu\,n_3(\theta).\\ \label{49}
\frac{\rd n_3}{\rd\theta}+\mu\,n_3&=2\,\sin\theta.
\end{align}
So,
\begin{equation}\label{50}
n_3(\theta)=\frac{2\,[\mu\,\sin\theta-\cos\theta+\exp(-\mu\,\theta)]}{1+\mu^2},
\end{equation}
resulting in
\begin{align}\label{51}
f_3(\theta)&=\frac{2\,\mu\,[\mu\,\sin\theta-\cos\theta+\exp(-\mu\,\theta)]}{1+\mu^2},\\ \label{52}
\tau_3(\theta)&=\frac{2\,\mu\,\sin\theta+(\mu^2-1)\,\cos\theta+2\,\exp(-\mu\,\theta)}{1+\mu^2},
\end{align}
from which
\begin{equation}\label{53}
\tau_3\left(\frac{\pi}{2}\right)=\frac{2\,[\mu+\,\exp(-\mu\,\pi/2)]}{1+\mu^2}.
\end{equation}
as it is expected, for $\mu\to 0$ these tend to the previous results.
For large $\mu$,
\begin{align}\label{54}
n_3(\theta)&=\frac{2\,\sin\theta}{\mu}+O(\mu^{-2}).\\ \label{55}
f_3(\theta)&=2\,\sin\theta+O(\mu^{-1}).\\ \label{56}
\tau_3(\theta)&=\cos\theta+O(\mu^{-1}).\\ \label{57}
\tau_3\left(\frac{\pi}{2}\right)&=\frac{2\,\sin\theta}{\mu}+O(\mu^{-2}).
\end{align}
\section{Concluding remarks}
A system was studied which consists of a massive uniform rope wrapped
around a fixed cylinder, with the rope having two vertical ends
supported by external forces. It was found that in order that 
the rope remain tightly wrapped around the cylinder, the supporting force 
should be larger than a minimum value. Then a situation was studied
where the rope looses its contact at some bottom part of the cylinder.
Conditions for such an equilibrium were investigated, and a relation was
found between the supporting force and the amount the ends of the rope fall.
Finally, the effect of the friction was taken into account and the system
was particularly studied at the threshold where the rope looses its contact 
with some part of the cylinder, for which the value of the supporting force 
was determined in terms of the friction coefficient.
\\[\baselineskip]
\textbf{Acknowledgement}:  This work was supported by
the research council of the Alzahra University.

\end{document}